\documentclass[preprint,authoryear,12pt]{elsarticle}

\usepackage{amssymb}

\usepackage{amsmath}
\usepackage{graphicx}
\usepackage{natbib}
\usepackage{color}
\usepackage{multirow}
\definecolor{Dark}{gray}{.90}
\def\mathbi#1{\textbf{\em #1}}

\journal{Medical Image Analysis}

\begin{document}

\begin{frontmatter}



\title{A blind deconvolution approach to recover effective connectivity brain networks from resting state fMRI data}


\author[a,b]{Guo-Rong Wu}
\author[c]{Wei Liao}
\author[d]{Sebastiano Stramaglia}
\author[b,e]{Ju-Rong Ding}
\author[b]{Huafu Chen}
\author[a]{Daniele Marinazzo\corref{cor1}}
\ead{daniele.marinazzo@ugent.be}

\address[a]{Faculty of Psychology and Educational Sciences, Department of Data Analysis, Ghent University, Henri Dunantlaan 1, B-9000 Ghent, Belgium.}
\address[b]{Key Laboratory for NeuroInformation of Ministry of Education, School of Life Science and Technology, University of Electronic Science and Technology of China, Chengdu 610054, China.}
\address[c]{Center for Cognition and Brain Disorders and the Affiliated Hospital, Hangzhou Normal University, Hangzhou 310015, China.}
\address[d]{Dipartimento di Fisica, Universit\'a degli Studi di Bari and INFN, via Orabona 4, 70126 Bari, Italy.}
\address[e]{Department of Psychological and Brain Sciences - Programs in Neuroscience and Cognitive Science, Indiana University, Bloomington, Indiana 47405, USA.}

\cortext[cor1]{Corresponding author}

\begin{abstract}
A great improvement to the insight on brain function that we can get from fMRI data can come from effective connectivity analysis, in which the flow of information between even remote brain regions is inferred by the parameters of a predictive dynamical model. As opposed to biologically inspired models, some techniques as Granger causality (GC) are purely data-driven and rely on statistical prediction and temporal precedence. While powerful and widely applicable, this approach could suffer from two main limitations when applied to BOLD fMRI data: confounding effect of hemodynamic response function (HRF) and conditioning to a large number of variables in presence of short time series. For task-related fMRI, neural population dynamics can be captured
by modeling signal dynamics with explicit exogenous inputs; for resting-state fMRI on the other hand, the absence of explicit inputs makes this task more difficult, unless relying on some specific prior physiological hypothesis. In order to overcome these issues and to allow a more general approach, here we present a simple and novel
blind-deconvolution technique for BOLD-fMRI signal. In a recent study it has been proposed that relevant information in resting-state fMRI can be obtained by inspecting the discrete events resulting in relatively large amplitude BOLD signal peaks. Following this idea, we consider resting fMRI as 'spontaneous event-related',
we individuate point processes corresponding to signal fluctuations with a given signature, extract a region-specific HRF and use it in deconvolution, after following an alignment procedure. Coming to the second limitation, a fully multivariate conditioning with short and noisy data leads to computational problems due to overfitting. Furthermore, conceptual issues arise in presence of redundancy. We thus apply partial conditioning to a limited subset of variables in the framework of information theory, as recently proposed. Mixing these two improvements we compare the differences between BOLD and deconvolved BOLD level effective networks and draw some conclusions. 
\end{abstract}


\end{frontmatter}


\section{Introduction}
	We can learn a lot on the functioning of the human brain in health and disease when we consider it as a large-scale complex network, whose properties can be analyzed using graph theoretical analysis \citep{bullmore2009complex}. With the advent of miscellaneous and noninvasive MRI techniques, this \textit{connectome} has been mainly characterized by either structural or functional connectivity.  Structural connectivity is commonly based on white matter tracts quantified by diffusion tractography \citep{hagmann2008mapping}; functional connectivity relies on the other hand on statistical dependencies such as temporal correlation \citep{salvador2005neurophysiological}. An important addition to this framework can come from effective connectivity analysis \citep{Stephan2012}, in which the flow of information between even remote brain regions is inferred by the parameters of a predictive dynamical model. \\
	
	For some techniques, such as dynamic causal modelling (DCM) and structural equation modelling \citep{buchel1997modulation,friston2003dynamic}, these models are built and validated from specific anatomical and physiological hypotheses. Other techniques such as Granger causality analysis (GCA) \citep{bressler2011wiener}, are on the other hand data-driven and rely purely on statistical prediction and temporal precedence. While powerful and widely applicable, this last approach could suffer from two main limitations when applied to blood-oxygenation level-dependent (BOLD)-functional MRI (fMRI) data: confounding effect of hemodynamic response function (HRF) and conditioning to a large number of variables in presence of short time series. Early interpretation of fMRI based directed connectivity by GCA always assumed homogeneous hemodynamic processes over the brain; several studies have pointed out that this is indeed not the case and that we are faced with variable HRF latency across physiological processes and distinct brain regions  \citep{roebroeck2009identification,valdes2011effective}. Recently, a number of studies have addressed this issue proposing to model the HRF according to several recipes \citep{bakhtiari2012subspace,havlicek2011dynamic,havlicek2010dynamic,ryali2011multivariate}. As well, a recent study has proposed that it would still feasible to infer connectivity at BOLD level, under the assumption that Granger causality is theoretically invariant under filtering \citep{Barnett2011404} and that the HRF can be considered as a filter. It is still unclear whether and how specific effects related to HRF disturb the inference of temporal precedence. In addition a simulated or experimental ground truth is difficult to obtain, though some studies on simulated fMRI data have tried to reveal the relationship between neural-level and BOLD-level causal influence \citep{deshpande2010effect,smith2011network}. A considerable help to obtain the HRF for deconvolution could come from multimodal imaging where the high temporal resolution of EEG is combined to the high spatial resolution of fMRI, but this experimental approach is still far from being optimal and widely applicable. HRF has been studied almost since the early days of fMRI \citep{handwerker2012continuing}. For task-related fMRI, neural population dynamics can be captured by modeling signal dynamics with explicit exogenous inputs \citep{friston2008variational,riera2004state}, i.e. deconvolution according to the explicit task design is possible in this case \citep{buxton1998dynamics,friston2000nonlinear,glover1999deconvolution}. For resting-state fMRI on the other hand, the absence of explicit inputs makes this task more difficult, unless relying on some specific prior physiological hypothesis \citep{friston2008variational,havlicek2011dynamic}. In order to overcome these issues and to allow a more general approach, here we present a simple and novel blind-deconvolution technique for BOLD-fMRI signal.\\
	
	Coming to the second limitation, in order to distinguish among direct and mediated influences in multivariate datasets it is necessary to condition the analysis to other variables. A bivariate analysis would indeed lead to the detection of many false positives. In presence of a large number of variable and short time series, a fully multivariate conditioning could lead to computational problems due to the overfitting. Furthermore, conceptual issues would arise in presence of redundant variables \citep{angelini2010redundant,marinazzo2010grouping}. In this paper we thus apply partial conditioning for Granger Causality (PCGC)\footnote{Please note that this approach is different from partial Granger causality (PGC) (Guo et al. (2008), Journal of Neuroscience Methods, {\bf 172}, 79)}
 to a limited subset of variables, as recently proposed \citep{marinazzo2011causal} for reconstructing the BOLD and deconvolved BOLD level effective connectivity network (ECN) and compare them.

\section{Materials and methods}

\subsection{Blind-deconvolution in resting-state fMRI data}
	Hemodynamic deconvolution of BOLD signal is performed as described in \citep{david2008identifying,glover1999deconvolution}. Under the assumption that the transformation from neural activation to BOLD response can be modeled as a linear and time invariant system, measured fMRI data $b(t)$ can be seen as the result of the convolution of neural states $s(t)$  with a HRF $h(t)$: 
\begin{equation}
b(t)=s(t)\otimes h(t)+\epsilon(t)
\end{equation}
Where $t$ is the time and $\otimes$ denotes convolution. 
$\epsilon (t)$ is the noise in the measurement, which we assume to be white. Since the right side of the above equation includes three unobservable quantities, in order to solve the equation for $h(t)$ we need to substitute $s(t)$ with a hypothetical model of the neural activation for $s(t)$. Here we employ a simple on-off model of activation to model $s(t)$:
\begin{equation}
\hat{s}(t)=\sum_{\tau=0}^{\infty}\delta(t-\tau) 
\end{equation}
where $\delta(t-\tau)$ is the delta function.
This allows to fit the HRF $h(t)$ according to $\hat{s}(t)$ using a canonical HRF (two gamma functions) and two derivatives (multivariate Taylor expansion: temporal derivative and dispersion derivative) \citep{friston2000nonlinear}, as is common in most fMRI studies.\\
Once calculated $h(t)$, we can obtain an approximation $\widetilde{s}(t)$ of the neural signal from the observed data  using a Wiener filter
\begin{equation}
\widetilde{s}(t)=d(t)\otimes b(t)
\end{equation}
Let $H(\omega)$, $B(\omega)$, $E(\omega)$, and $D(\omega)$ be the Fourier transforms of $h(t)$, $b(t)$, $\epsilon(t)$, and $d(t)$, respectively. Then
\begin{equation}
D(\omega)= \frac{H^*(\omega)}{\left| H(\omega)\right|^2 + \left| E(\omega)\right|^2},
\end{equation}
where $*$ denotes complex conjugate. The estimation $\widetilde{s}(t)$ of the neural states $s(t)$ is then given by
\begin{equation}
\widetilde{s}(t)=FT^{-1} \left\lbrace D(\omega)B(\omega)\right\rbrace= FT^{-1} \left\lbrace \frac{H^*(\omega)B(\omega)}{\left| H(\omega)\right|^2 + \left| E(\omega)\right|^2}\right\rbrace.
\end{equation}
Where $FT^{-1}$ is the inverse Fourier transform operator.\\

	For task-related fMRI, the stimulus function provides the prior expectations about neural activity and a generative model whose inversion corresponds to deconvolution; this is in principle not the case for resting-state fMRI. Nonetheless there is increasing evidence of specific events and neural states that govern the dynamics of the brain at rest \citep{deco2012ongoing,petridou2012periods}. Furthermore, Tagliazucchi et al. proposed that these events are reflected by relatively large amplitude BOLD signal peaks and thus that such fluctuations could encode relevant information from resting-state fMRI recordings \citep{tagliazucchi2012criticality}. Inspired by their work, we consider resting-state fMRI as \textit{spontaneous event-related}, and we propose to extract the HRF from those pseudo-events. After doing this, we can employ the deconvolution model in the same way as described above. It is known that the BOLD response is much slower than the neural activation that is presumed to drive it. Consequently, the peak of the BOLD signal lags behind the peak of neural activation (i.e. by $\kappa$ points). So here we assume that these events are generated from $\hat{s}(t)$.\\
	
	Glover pointed out that the noise spectrum in task-related fMRI can be obtained from time series measurements in nonactivated cortical regions \citep{glover1999deconvolution}; here we extend the model to cope with resting-state fMRI for which there is no explicit activation. In this study we assumed covariance of noise $\epsilon$ equal to $cov\left[b(t)-\hat{s}(t)\otimes h(t)\right]$.\\
	
\begin{figure}		
\begin{center}
\includegraphics[scale=0.8]{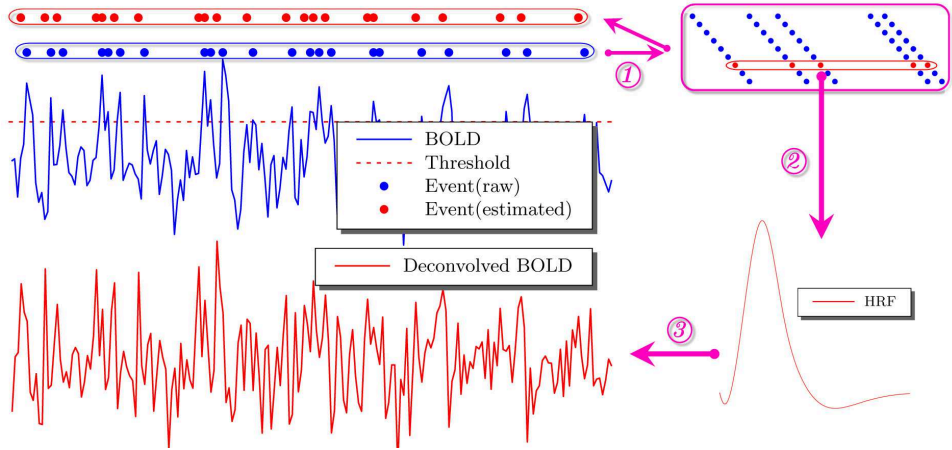}
\end{center}
\caption{Flow chart for blind-deconvolution procedure. 1. the pre-processed (detrended and normalized) observed BOLD signal is evaluated against a given threshold obtaining several sets of putative onsets for pseudo-events. 2. the time deviation of the timing sets is adjusted; the set with smallest noise error covariance will represent the event. 3. the observed BOLD signal is deconvolved into a neural signal by using the corresponding HRF.\label{fig1}}
\end{figure}

	In order to obtain a value for $\kappa$, we search all integer values in the interval $[0 \: \kappa_{max}]$, where $\kappa_{max}$ is an arbitrary maximum value, choosing the one for which the noise error covariance is smallest as the onset. By this method we can perform deconvolution on all BOLD signals, requiring no information on timing or a priori spatial information of events; furthermore, the time series could be the average of time series over a region of interest with any scale, or series extracted by independent or principal component analysis. A flow chart for BOLD signal deconvolution is shown in Fig.\ref{fig1}.\\
	This is the pseudo-code for our procedure.\\
	\vspace{.5cm}
\fbox{
\colorbox{Dark}{
\begin{minipage}[h]{0.9\textwidth}
Pseudo-code

\begin{itemize}
\item[i] Preprocess time series (e.g. detrend, normalize etc.).
\item[ii] Find a time set $S$ in which the BOLD values exceed a given threshold around a local maximum.
\item[iii]
choose a maximum time delay $\kappa_{max}$ \\
FOR $n=0$ to $\kappa_{max}$ \\

 $S_n=S-n$\\

 $\hat{s}_n(t)=1, t\in S_n; \hat{s}_n(t)=0, t\notin S_n $.\\

 Fit a general linear model using $\hat{s}_n$ and canonical HRF with time and dispersion derivatives.\\

END FOR
\item[iv] Let $\epsilon_\kappa = \min\limits_{0\leq n \leq \kappa_{max}} \{\epsilon_n\}$, where $\epsilon_n = cov\left[b(t)-\hat{s}_n(t)\otimes h_n(t)\right]$.
\item[v] Follow equation 4 and 5, using HRF $h_\kappa$ and $\epsilon_\kappa$ for deconvolution, get $\widetilde{s}(t)$.
\end{itemize}
\end{minipage}
}
}

\subsection{Partially conditioned Granger causality} 

	Here we employ a methodology proposed in \citep{marinazzo2011causal} which allows to compute Granger causality conditioned to a limited number of variables in the framework of information theory. The idea is that conditioning on a small number of variables, chosen as the most informative for the candidate driver variable, is sufficient to remove indirect interactions for sparse connectivity patterns. \\
	
	We consider $n$  covariance-stationary variables $\left\lbrace x_i(t)\right\rbrace _{i=1,\cdots,n}$, denoting the state vectors as:
\begin{equation}
X_{\alpha}(t)=\left( x_{\alpha}(t-m),\cdots , x_{\alpha}(t-1) \right)
\end{equation}
$m$ being the model order. Let $\epsilon(x_{\alpha}|Y)$ be the mean squared error prediction of $x_{\alpha}$ on the basis of the vectors $Y$. The partially conditioned Granger causality index  $c(\beta \rightarrow \alpha)$ is defined as follows:
\begin{equation}
c(\beta \rightarrow \alpha)=
log\frac{\epsilon \left( x_{\alpha} | \mathbi{Z}\right)}{\epsilon \left( x_{\alpha} | \mathbi{Z} \cup X_{\beta}\right)}
\end{equation}
\\

	Where  $\mathbi{Z}=\left\lbrace X_{i_1},\cdots,X_{i_{n_d}} \right\rbrace$ is a set of the $n_d$ variables, in $\{X_1,X_2,\cdots, X_n\}\backslash X_{\beta}$, which are most informative for $X_{\beta}$. We adopt the following approximate strategy for $\mathbi{Z}$ : given the previous $Z_{k-1}$, the set $Z_k$ is obtained adding the variable with greatest information gain. This is repeated until $n_d$ variables are selected.

\subsection{Simulation Datasets: NetSim}

	 A method for establishing a ground truth for fMRI data has not reached a general consensus. Recently a benchmark dataset, NetSim \citep{smith2011network} has attracted a lot of attention. Previous studies have shown that lag-based methods perform very poorly on these datasets; it is anyway worthy to mention that these data are simulated under the DCM framework, contain no reciprocal connections and only Gaussian noise, limiting their universality as ground truth. Here we analyzed the largest of these datasets, consisting of 50 nodes. After deconvolution the sensitivity improved significantly, increasing from 20$\%$ to 30$\%$. Also the specificity improved from 88$\%$ to 94$\%$. This does not render GC the method of choice for these data, for which we also have to point out that ''neural events'' and noise are not distinguishable, but gives nonetheless an indicative result for the usefulness of deconvolution of the BOLD signal.

\subsection{Resting-State fMRI Datasets}
	In order to investigate the role of repetition time (TR) on the deconvolution procedure and on the effective network reconstruction, our analyses were performed on a resting-state fMRI dataset which has been publicly released in the ''1000 Functional Connectomes Project''\footnote{http://fcon\_1000.projects.nitrc.org, accessed march 2012.\label{f2}}. All participants had no history of neurological and psychiatric disorders and all gave the informed consent approved by local Institutional Review Board. During the scanning participants were instructed to keep their eyes closed, not to think of anything in particular, and to avoid falling asleep.\\

	Two data sets with different TR (TR=1.4s and TR=0.645s) were acquired on Siemens 3T Trio Tim scanners using developed multiplexed echo planar imaging \citep{feinberg2010multiplexed}. As specified in detail below, two resting-state fMRI data are included in the protocol - a TR=0.645s (3mm isotropic voxels, 10 minutes) to provide optimal temporal resolution and TR=1.4s (2mm isotropic voxels, 10 minutes) to provide optimal spatial resolution. The third data set, acquired on a 4T scanner, contains standard resting-state fMRI acquisitions with a longer TR (TR=3, 4mm isotropic voxels, 5 minutes). For more detail on subject and data information, please see website$^{\ref{f2},}$ \footnote{http://fcon\_1000.projects.nitrc.org/indi/pro/eNKI\_RS\_TRT/FrontPage.html.}.

\subsection{Data preprocessing}
	Preprocessing of resting-state images was performed using the Statistical Parametric Mapping software (SPM8, http://www.fil.ion.ucl.ac.uk/spm). The preprocessing included slice-timing correction relative to middle axial slice for the temporal difference in acquisition among different slices, head motion correction, spatial normalization into the Montreal Neurological Institute stereotaxic space, resampling  to 3-mm isotropic voxels. 8(9) subjects were excluded from the dataset with TR=0.645s (TR=1.4s) because either translation or rotation exceeded $\pm 1.5$ mm or $\pm 1.5^\circ$, resulting in 16(TR=0.645s) and 15(TR=1.4s) subjects each one scanned in two sessions which were used in the analysis). One subject whose data were too noisy was excluded from the TR=3 dataset, resulting in 10 subjects used in the analysis. In order to avoid introducing artificial local spatial correlations between voxels, no spatial smoothing was applied for further analysis, as previously suggested \citep{zhang2011altered,salvador2005neurophysiological,liao2011small}.

\subsection{Anatomical parcellation and analysis}
	The functional images were segmented into 90 regions of interest (ROI) using automated anatomical labeling (AAL) template as reported in previous studies. For each subject, the representative time series of each ROI was obtained by averaging the fMRI time series across all voxels in the ROI \citep{salvador2005neurophysiological}. Several procedures were used to remove possible spurious variances from the data through linear regression. These were i) six head motion parameters obtained in the realigning step, ii) signal from a region in cerebrospinal fluid, iii) signal from a region centered in the white matter, iv) global signal averaged over the whole brain. The BOLD time series were deconvolved into neural state signal using the above mentioned approach. 

\subsection{Effective connectivity network analysis}
	The topological properties of the effective connectivity network were defined on the basis of a $90\times 90$ binary directed graph $G$, consisting of nodes and directed edges:
\begin{equation}
eij = \left\{ {\begin{array}{*{5}c}
1,  F_{i\rightarrow j} >T;  \\
0,  otherwise \\
\end{array}} \right.
\end{equation}
Where $e_{ij}$ refers to the directed edge from ROI $i$ to ROI $j$ in the graph. $T$ indicates the threshold. In a directed graph $e_{ij}$ is not necessarily equal to $e_{ji}$. Considering that the graph we focused on is directed, all topological properties were calculated on incoming and outgoing matrix, respectively. Graph theoretical analyses were carried out on the effective connectivity network using the Brain Connectivity Toolbox \citep{rubinov2010complex}.

\subsection{Threshold selection}
	As previous studies suggested that the brain networks of each subject normally differ in both the number and weighting of the edges \citep{zhang2011altered,liao2011small}, we applied a matching strategy to characterize the properties of effective connectivity network. Both the global and local network efficiencies have a propensity for being higher with greater numbers of edges in the graph \citep{wen2011discrete}. Modifying the sparsity values (number of edges) of the adjacency matrix also altered the graph's structure. As a consequence it was suggested that the graphs to be compared must have (a) the same number of nodes and (b) the same number of edges \citep{bullmore2011brain}.	The cost was defined as the ratio of the number of existing edges divided by the maximum possible number of edges in a network. Since there is currently no formal consensus regarding selection of cost thresholds, here we selected a range of 0.05 to 0.14 with step = 0.01 for subsequent network analyses. The lower bound was chosen as the one yielding a sparse graph with mean degree $\geq 2 ln(90)$ (total number of edges $\geq 405$ where $405/90^2=0.05$). The upper threshold corresponded to the smallest significant value of Granger causality (F-test with $p=0.05$) across all subjects).

\subsection{Network metrics}
	For effective connectivity network at each cost threshold, we calculated both overall topological properties and nodal characteristics \citep{rubinov2010complex}. The overall topological properties included i) small-worldness ($\sigma$), related to normalized clustering coefficient ($\gamma$) and normalized characteristic path length ($\lambda$); ii) network efficiency, divided in local efficiency ($E_{loc}$) and global efficiency ($E_{glob}$). The nodal characteristics included i) the nodal degree, that quantifies the extent to which a node is relevant to the graph, and ii) the nodal efficiency, that quantifies the importance of the nodes for the communication within the network \citep{bassett2006small}. Furthermore we calculated the area under the curve (AUC) across all cost values for the above mentioned network properties. This quantity represents a summarized scalar for topological characterization of brain networks independent of single cost threshold selection.

\section{Results}

\subsection{Reconstruction of HRF}
\begin{figure}
\begin{center}
\includegraphics[scale=0.6]{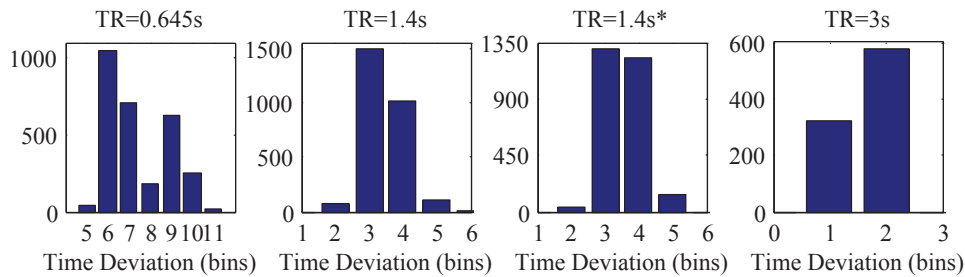}
\caption{Histogram of the values of time deviation $\kappa$. ($*$:without regression of global signal)
\label{fig2}}
\end{center}
\end{figure}

	We tested the proposed deconvolution method on resting-state fMRI data; following the procedure summarized in the box, firstly we set a maximum time lag from a given threshold crossing, and obtain an optimal value for this lag, denoted with $\kappa$. The histograms for $\kappa$, reported in Fig.\ref{fig2} show a maximum around $4 \sim 6$s , which is consistent with a previous study according to which the latency delay is $4 \sim 8$s in gray matter \citep{lee1995discrimination}. It is worth to mention that the lower TR could allow a more accurate estimation of the lag.\\
	
\begin{figure}
\begin{center}
\includegraphics[scale=1]{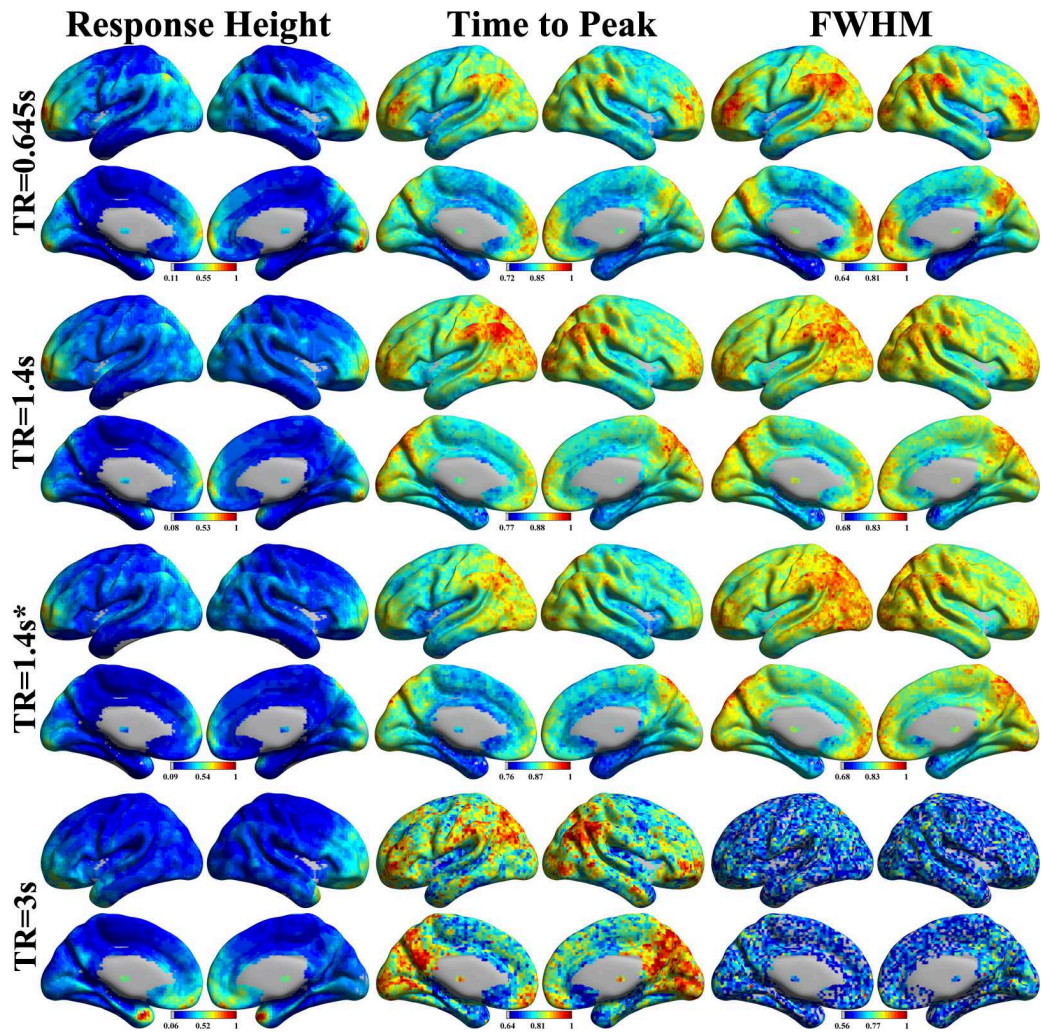}
\caption{Spatial distribution of HRF shape: response height, time to peak and full-width at half-max (all values have been normalized, keeping range from 0 to 1).($*$:without regression of global signal)\label{fig3}}
\end{center}
\end{figure}

	To assess the effect of deconvolution, we compared the shape of voxel based HRF over the whole brain using different TRs. We focused on three parameters: response height, time-to-peak, and full-width at half-max (FWHM) as potential measures of response magnitude, latency, and duration. Using principal component analysis we determined the average intersubject variability of HRF maps. We found that the first component of HRF accounted for $81.7\pm 2.9\%$(response height), $98.1\pm 1.2\%$(time to peak) and $95.6\pm 3.5\%$(FWHM) of the variance. Furthermore, the spatial distribution is very similar to the mean group map. The mean group results are plotted in Fig.\ref{fig3}.
	 The response height, time to peak and FWHM of HRFs differ across brain regions, as a consequence of multiple factors including neural activity differences, global magnetic susceptibilities, vascular differences, baseline cerebral blood flow, slice timing differences etc. \citep{handwerker2004variation}. These patterns are remarkably similar across subjects and TRs, reflecting the robustness of the proposed approach.\\
	
\subsection{Variance stability of causality matrix}
\begin{figure}
\begin{center}
\includegraphics[scale=0.6]{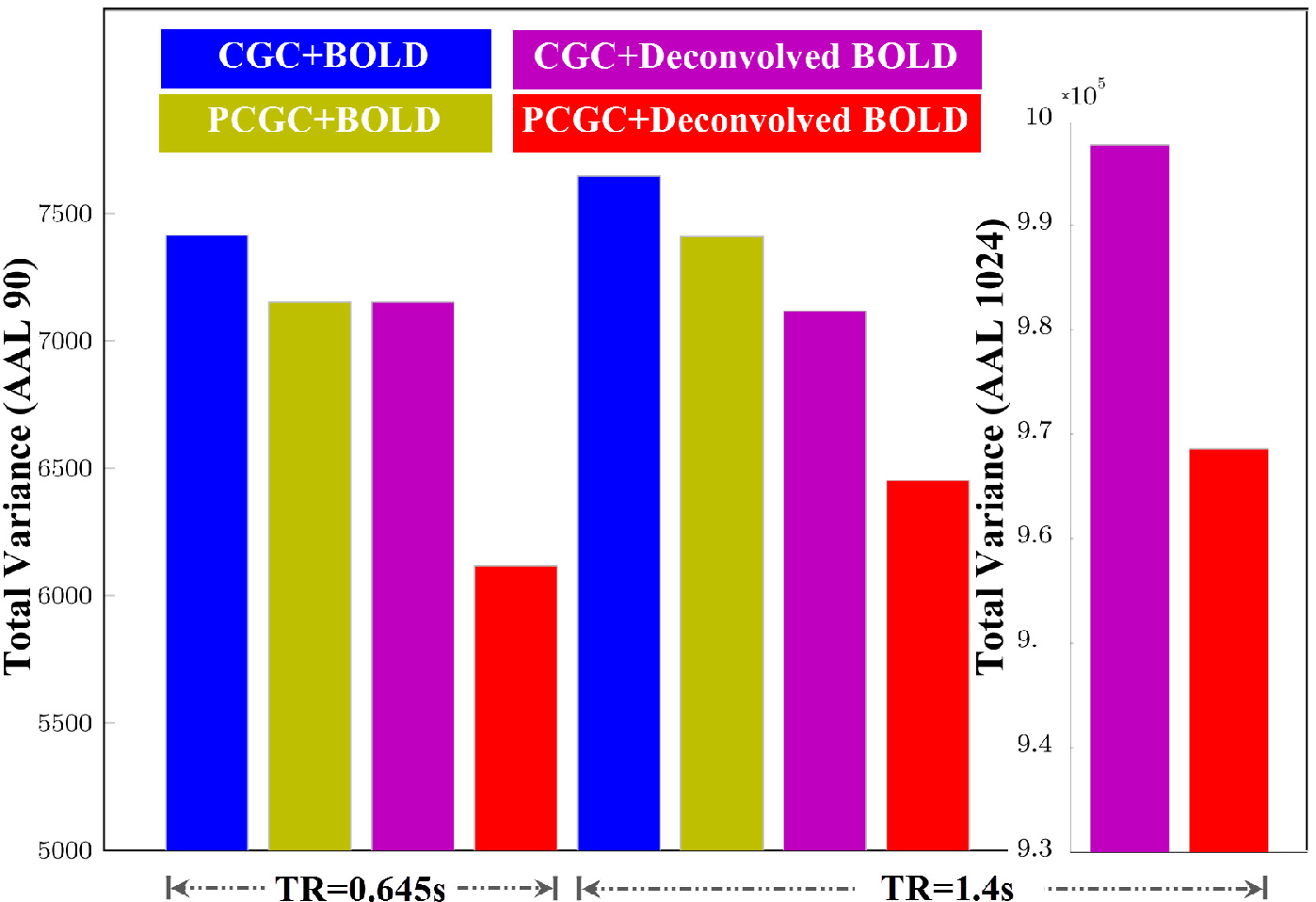}
\caption{Total variance of causality matrix across all subjects. Full conditional Granger causality (CGC) and PCGC combined with BOLD and deconvolved BOLD level signal were used for construction of causality matrix. \label{fig6}}
\end{center}
\end{figure}

	As another indicator of the stability of the proposed joint of deconvolution and PCGC approach we tested the variance of causality matrix across all subjects. We calculated the variance of the Granger causality matrix obtained both on the BOLD and deconvolved BOLD level signal. Firstly, we converted the matrix to Z-scores, then we calculated the variance of each matrix element, finally summing up the all these values into an overall variance index. The variance of Granger causality matrix obtained from the deconvolved signal is much lower than the one of the BOLD level matrix for all TR values (Fig.\ref{fig6}). Also, PCGC method kept the variance lower than full conditioned GC method. This result was confirmed testing a network at another scale using 1024 nodes (Fig.\ref{fig6}, the native AAL segmentation was parcellated into 1024 micro regions of interest of approximately identical size across both hemispheres \citep{zhang2011altered}; in this case we could not test full conditioned GC due to small number of samples).

\subsection{Global signal regression}
	As shown in previous studies, several sources of spurious variance should be removed by regression: motion artifacts, white matter and ventricular time courses. Still, the effects of regression against the global signal, calculated by averaging across all voxels within a whole brain mask, are debated. In order to evaluate this effect on our data we calculated spatial correlation between the the group mean image of HRF(response height, time to peak, FWHM) with and without regression of global signal in the preprocessing step, obtaining high Pearson correlation between them: r=0.97(response height), 0.90(time to peak), 0.88(FWHM). We can thus conclude that regression against global signal still preserved the spatial distribution.

\subsection{Effective connectivity network recovery with partial conditioning}
	When trying to reconstruct effective connectivity networks, we are faced with the problem of coping with a large number of variables, when the application of multivariate Granger causality may be questionable or even unfeasible, whilst bivariate Granger causality would detect also indirect interactions. Conditioning on a large number of variables requires an high number of samples in order to get reliable results. Reducing the number of variables that one has to condition over would thus provide better results for small data-sets. In the general formulation of Granger causality, one has no way to choose this reduced set of variables; on the other hand, in the framework of information theory, it is possible to individuate the most informative variables one by one.\\
	
\begin{figure}
\begin{center}
\includegraphics[scale=1]{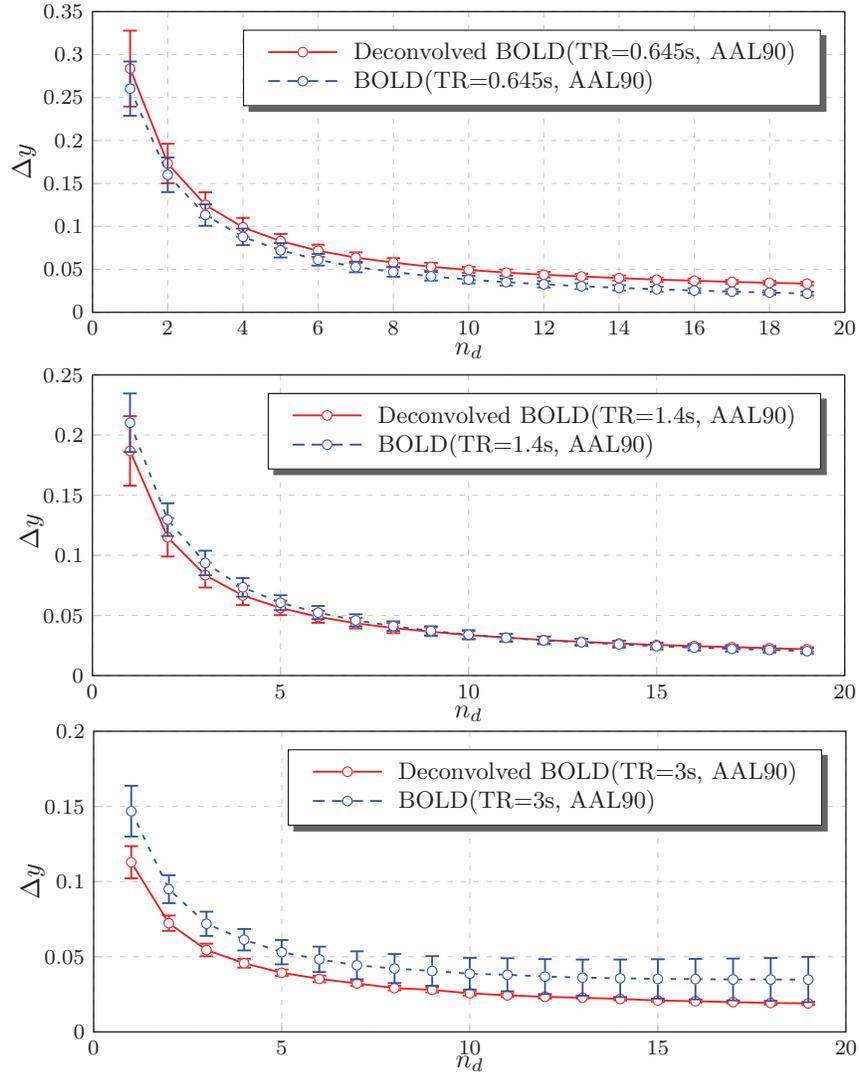}
\caption{The mutual information gain ($\Delta y$), when the $(n_d+1)$-th variable is included, is plotted versus $n_d$. The information gain is averaged over all the variables. 
\label{fig4}}
\end{center}
\end{figure}

	The optimal model order $m$ (order of the autoregressive model in Granger causality, embedding dimension in transfer entropy) for deconvolved BOLD and BOLD signal was determined by leave-one-out cross-validation, and was found to be 3 for TR=0.645s, 2 for TR=1.4s and 1 for TR=3s. Under the Gaussian assumption, we constructed effective connectivity network using PCGC method. We firstly have to determine the number of variables upon which conditioning. To do this we look at how much uncertainty we eliminate adding an extra variable, letting the number of conditioning variables included $n_d$ vary from 1 to 20. This uncertainty can be expressed in terms of the information that we gain adding an extra variable. In Fig.\ref{fig4}, we plot the information gain as a function of $n_d$; as expected, both this quantity and its increment decrease monotonically with $n_d$.  \\

	We can observe that the knee of the curves occurs when six variables are considered. This happens also when we consider different brain prior templates with 17 or 160 nodes (results not reported here). This could be connected to the fact that the average number of modules which explain the equal-time correlations of resting brain is close to six \citep{marinazzo2010grouping,salvador2005neurophysiological}, therefore picking one variable from each module is sufficient to have most of the information, about a given channel, that can be obtained from the remaining channels, and this independently on the number of nodes. The effect of deconvolution, for the information gain, is to qualitatively raise the curve for TR=0.645s, and to lower them for TR=1.4s. This trend (not statistically significant) might be the result of two competing effects, the fact the deconvolution may remove spurious correlations and/or restore genuine correlations obscured by noise.\\
	
	Synthesizing the knee of the curves, sensitivity and specificity, we consider $n_d=10$ as the most appropriate number of most informative variables to include in the conditioning procedure.\\

\subsection{Global characteristics of ECN}
	The global topological properties of brain ECN at deconvolved BOLD and BOLD level rely on the choice of thresholds. We used multiple cost thresholds and the AUC to evaluate the stability of the topological organization (Table 1). An higher number of differences between the two networks was found with a (relatively) longer TR (TR=1.4s). Specifically, the AUC of small-worldness ($\sigma$), normalized clustering coefficient ($\gamma$), clustering coefficient($C_p$) and local efficiency($E_{loc}$ ) displayed the most significant differences, similar to what emerged with TR=0.645s. For the data set with shorter TR we found significant differences in the characteristic path length and global efficiency of the outgoing network, whereas the most relevant differences were found for the incoming network with the longer TR.

\begin{table}
\begin{tabular}{|c|c|c|c|c|c|c|}
\hline \hline
\multirow{3}{*}{Global network parameter} &  
\multicolumn{4}{|c|}{AUC difference} \\
\cline{2-5}
 & \multicolumn{2}{|c|}{TR=0.645s} & \multicolumn{2}{|c|}{TR=1.4s}\\
\cline{2-5}

& Incoming & Outgoing & Incoming & Outgoing  \\
\hline
Sigma($\sigma$)  & Y & N & Y & Y  \\
\hline
Lambda($\lambda$)  & N & N & N & N \\
\hline
Gamma($\gamma$)  & Y & N & Y & Y \\
\hline
Characteristic path length($L_p$)  & N & Y & Y & N \\
\hline
Clustering coefficient($C_p$)  & Y & N & Y & Y \\
\hline
Global efficiency($E_{glob}$)  & N & Y & Y & N \\
\hline
Local efficiency($E_{loc}$)  & Y & N & Y & Y \\
\hline
\hline
\end{tabular}
\\
\caption{Comparison of AUC between deconvolved BOLD and BOLD.
$n_d=10$, Y: $p<0.05$, FDR corrected; N: otherwise.}
\end{table}

\subsection{Nodal characteristics of ECN}
\begin{figure}
\begin{center}
\includegraphics[scale=0.6]{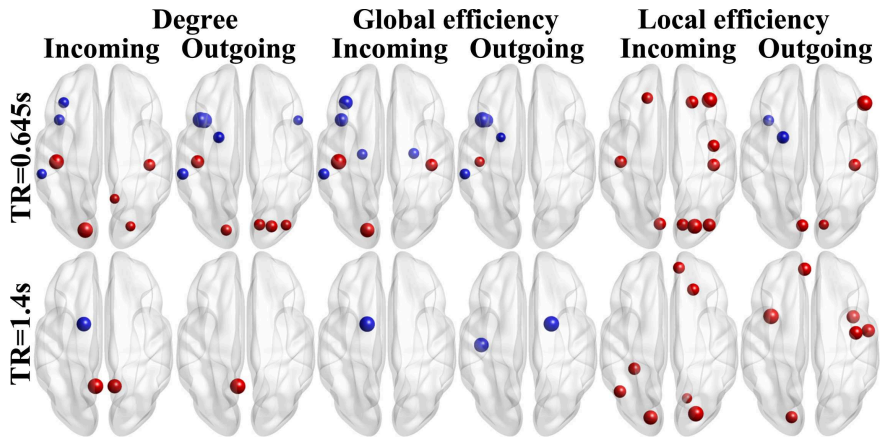}
\caption{Z-scores for Area under Curve from regional nodal parameters (deconvolved BOLD vs BOLD), $p<0.05$, FDR corrected, ($n_d=10$). Blue indicate negative values, red positive values. The point size is proportional to the absolute Z value.\label{fig5}}
\end{center}
\end{figure}

	Comparing the two networks on nodal degree, nodal global efficiency and nodal local efficiency revealed modifications in deconvolved BOLD and BOLD level (Fig.\ref{fig5}). The patterns of nodal degree modifications resembled to those of nodal global efficiency in incoming network in both TR=0.645s and TR=1.4s fMRI data sets. In addition, more brain regions showed modified nodal degree and (global/local) efficiency in  TR=0.645s rather than TR=1.4s data.	

\section{Discussion}
In this study we proposed a novel methodology to achieve deconvolution in resting state data using spontaneous pseudo events, and to apply partially conditioned Granger Causality to the analysis of fMRI data. In our opinion this joint approach is the most convenient to infer effective connectivity with Granger Causality from resting state fMRI data.
	
	In the absence of a well defined ground truth, and in the light of the still active and unresolved debate on the usefulness of HRF deconvolution Granger causality based connectivity, we limit ourselves to validate the stability of the proposed method and indicate a possible path for the continuation of this debate, quantifying and comparing the overall topological properties of large-scale ECNs on deconvolved BOLD-level versus BOLD-level signals, investigating also the effect of different time resolutions (TR=0.645s and TR=1.4s).\\
	
	Previous discussions on evaluating effective connectivity from fMRI data reached the conclusion that it is better to use state-space model for inferring causality on hidden neural states \citep{valdes2011effective,ryali2011multivariate,bakhtiari2012subspace}. A pioneering EEG-fMRI study provided the first experimental substantiation of the theoretical possibility to improve interregional coupling estimation from hidden neural states of fMRI \citep{david2008identifying}. Though promising \citep{friston2009causal}, these implications are still limited by the fact that multimodal recording is invasive and not applicable to healthy controls. As a consequence, data-driven methods for substantiating the confounding variability of haemodynamics have been developed. The two available types of state space models in estimation of HRF \citep{valdes2011effective}: the generic (linear canonical/spline HRF) \citep{glover1999deconvolution,marrelec2003robust} and biophysically informed models (DCM nonlinear HRF)\citep{friston2000nonlinear}. Generic models are widely applicable but lack specific biophysical constraints \citep{glover1999deconvolution,marrelec2003robust}, while biophysically informed models are constrained by the hypothesis itself \citep{friston2000nonlinear}. A recently proposed, biophysically informed bind deconvolution approach based on the state-of-the-art Cubature Kalman filtering could be a useful tool for resting-state fMRI \citep{havlicek2011dynamic}. In the present study, however, we use a simpler approach which employs the generic linear canonical HRF for deconvolution.  It is worth to point out that the significant differences between BOLD- and deconvolved BOLD-level effective connectivity found in complex network measures cannot absolutely exclude the misestimation of HRF. Furthermore HRF latency effect does not always critically affect the evaluation of mutual influence, so ECNs on BOLD and deconvolved BOLD level could have important consistencies \citep{supekar2012developmental}.\\	
	
Findings from brain connectivity studies have now demonstrated that the human brain network exhibits robust small-world topological properties, not only in the anatomical connectivity (reconstructed by diffusion tractography) \citep{hagmann2008mapping} and functional connectivity network \citep{salvador2005neurophysiological}, but also in effective connectivity network \citep{liao2011small}. The current results also suggested that the ECNs obtained from BOLD and deconvolved data, with shorter and longer TR, have prominent small-world attributes, which would thus be confirmed as a general signature of robust organization of complex brain networks. Small-worldness indicates indeed an optimal balance between segregated and integrated organization to process the information \citep{bassett2006small}. For relatively longer TR we found significant differences between BOLD and deconvolved ECNs. Although an explanation based on precise neurobiological mechanisms is still not evident, we can suggest that the BOLD effect results from a more complex sequence of effects linking neuronal activity, vascular changes and MRI signal \citep{logothetis2008we}. Hemodynamic delay, and hence the correct onset of the events is indeed hard to capture with a long TR \citep{laufs2008recent}. \\

In complex networks organization, the normalized clustering coefficient  and the clustering coefficient are two key measures. They quantify the extent of local cliquishness or of local efficiency of information transfer of a network \citep{bullmore2011brain}, reflecting the local properties of network topologies. For longer TR, we observed significant differences between the two level ECNs. Thus the short-scale or local-scale network properties are indeed affected by deconvolution. Moreover, the normalized characteristic path length and the characteristic path length quantify global efficiency or the capability for parallel information propagation of a network \citep{bullmore2009complex}. These two measurements along with global efficiency are mainly associated with long-range connections ensuring effective interactions or rapid transfers of information \citep{he2009impaired}. It is widely accepted that long-range axonal connectivity being an important indicator of the functional{-}anatomical organization of the human cortex \citep{knosche2011role}. This study reported no differences in long-range network organization. \\

	It is known that resting-state functional connectivity studies using either seed functional connectivity or independent component analysis benefit from higher sampling rates to adequately sample undesirable respiration and cardiac effects \citep{birn2008effect}, while for event-related fMRI, faster sampling could allow for a better characterization of the hemodynamic response. The same applies to GCA. The previous simulations showed that accuracy of Granger causality depends on volume TR, faster sampling interval increased the detection capacity of GCA of fMRI data to neural causality \citep{deshpande2010effect,roebroeck2005mapping}. In this paper, we focus on resting-state fMRI data with TR=0.645s and 1.4s to maximally escape information loss due to low sampling. Considering the limitation of acquisition sequence, the conventional fast TR data acquisition brings to the loss of the fine spatial resolution \citep{huettel2004functional,kim1994potential}. \\

	Other methodological considerations are worth to be mentioned. The first one concerns data preprocessing. As a  general idea spatial smoothing can reduce the noise and increase signal-to-noise ratio, therefore improving the accuracy of detecting of neural event \citep{huettel2004functional}. Here we do not include this step. As we used AAL template, spatial smoothing would blur the boundary among these regions, which may affect the GC inference. Temporal filtering is frequently a necessary step for functional connectivity analysis of resting-state fMRI data. In line with previous studies that considered a low model order in GCA \citep{hamilton2010investigating,liao2011small},  we did not performed low-pass filtering.\\

	Secondly, graph theoretic approach is one of the most powerful and flexible approaches to investigate functional and structural brain connectome; still some controversies remain, concerning  the definition of network nodes and edges \citep{bullmore2011brain,wig2011concepts}. Different node definitions by prior anatomic brain templates \citep{wang2009parcellation} or node scales \citep{fornito2010network,zalesky2010whole} could produce different results. In future works, more brain templates and more node scales comparison for effective connectivity network should be explored. \\
	
\section*{Acknowledgments}
H. Chen was supported by the Natural Science Foundation of China (No. 61125304 and No. 61035006). G.R. Wu gratefully acknowledges the financial support from China Scholarship Council(2011607033).

\section*{References}








\end{document}